\documentclass[12pt, twocolumn, english]{iopart}
\expandafter\let\csname equation*\endcsname\relax
\expandafter\let\csname endequation*\endcsname\relax
\usepackage{amsmath}
\usepackage[T1]{fontenc}
\usepackage[english]{babel}
\usepackage{bm}
\usepackage{bbm}
\usepackage{iopams}
\usepackage{amssymb}
\usepackage{bbold}
\usepackage{graphicx}
\usepackage{appendix}
\usepackage{caption,booktabs}
\usepackage{array}
\newcolumntype{P}[1]{>{\centering\arraybackslash}p{#1}}

\usepackage{color}

\usepackage[%
  breaklinks=true,
  colorlinks=true,
  linkcolor=blue,
  urlcolor=blue,
  citecolor=blue
]{hyperref}
 \usepackage[numbers,sort&compress]{natbib}
 \usepackage{hypernat}
 \usepackage{pifont}% http://ctan.org/pkg/pifont

\begin{document}

\title{Signatures of magnetic inertial dynamics in two-sublattice ferromagnets}

\author{Ritwik Mondal$^{1,2}$}
\address{$^1$Department of Spintronics and Nanoelectronics, Institute of Physics ASCR, v.v.i., Cukrovarnick\'a 10, Prague 6, 162 53, Czech Republic}
\address{$^2$Department of Physics and Astronomy, Uppsala University, Box 516, SE-75120 Uppsala, Sweden}

\ead{mondal@fzu.cz}

%----------------------------------------------------------------------
\begin{abstract}
The magnetic inertial dynamics have been investigated for one sublattice ferromagnets. Here, we develop the magnetization dynamics in two-sublattice ferromagnets including the intra- and inter-sublattice inertial dynamics. First, we derive the magnetic susceptibility of such a ferromagnet. Next, by finding the poles of the susceptibility, we calculate the precession and nutation resonance frequencies. Our results suggest that while the resonance frequencies show decreasing behavior with the increasing intra-sublattice relaxation time, the effect of inter-sublattice inertial dynamics is contrasting.        

\end{abstract}
%----------------------------------------------------------------------

\section{Introduction}
Ultrafast manipulation of electrons' spin remains at the heart of future generation spin-based memory technology \cite{bigot09,Stanciu2007,Kimel_2007_JPCM_review}. It has been observed that a fs laser pulse is capable of demagnetizing a ferromagnetic material~\cite{Bigot1996,Koopmans2000,Koopmans2003JPCM}. On the other hand, using these ultrashort pulses, magnetic switching has been reported in ferrimagnetic~\cite{Mangin2014,Hassdenteyfel2013,Kimel2005} and ferromagnetic materials~\cite{lambert14,John2017}. These observations have been explained through the spin dynamics within Landau-Lifshitz-Gilbert (LLG) equation of motion~\cite{Ostler2012,Wienholdt2013,Gerlach2017,Frietsch2020}.

\begin{figure}[hbt!]
    \centering
    \includegraphics[scale = 1]{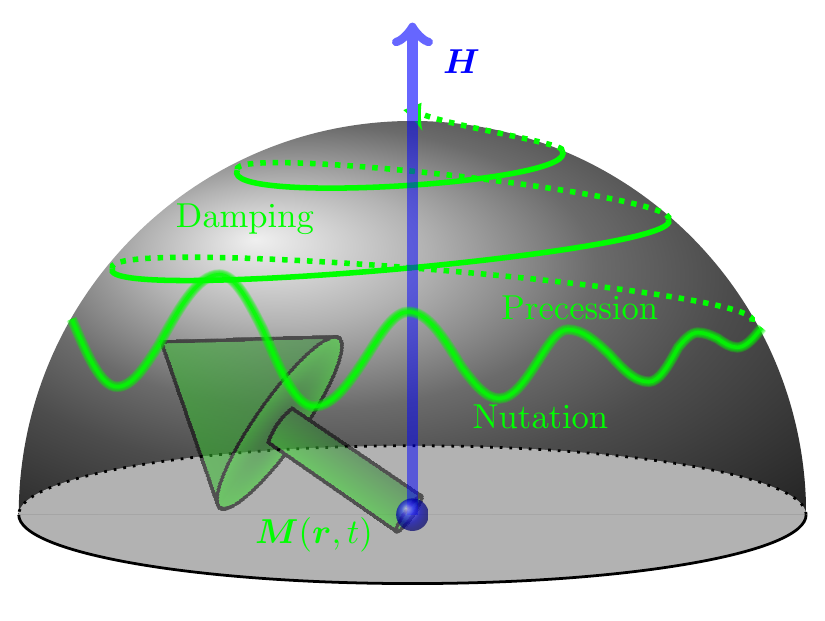}
    \caption{ Schematic depiction of ILLG equation of motion. }
    \label{Fig1}
\end{figure}

The phenomenological LLG spin dynamics consists of spin precession and a transverse damping~\cite{landau35,Gilbert1955,Gilbert2004}. Such an equation of motion has been derived from a relativistic Dirac theory, where the transverse damping is found to originate from spin-orbit coupling~\cite{hickey09,Mondal2015a,Mondal2016,Mondal2018PRB}. However, at ultrashort timescales, the traditional LLG equation needs to be supplemented by several other spin torque terms~\cite{Mondal2019PRB}. Especially, at the ultrafast timescales, the magnetic inertia becomes particularly relevant~\cite{Henk2012}. The effect of magnetic inertia has been incorporated within extended LLG dynamics as a torque due to the second-order time derivative of the magnetization $\bm{M}(\bm{r},t)$. The inertial LLG (ILLG) equation of motion reads~\cite{Ciornei2010thesis,Ciornei2011,Bhattacharjee2012}
\begin{align}
    \frac{\partial \bm{M}}{\partial t} & = \bm{M} \times \left[-\gamma \bm{H} +  \frac{\alpha}{M_0}\frac{\partial \bm{M}}{\partial t} + \frac{\eta}{M_0} \frac{\partial^2 \bm{M}}{\partial t^2}\right]\,,
    \label{ILLG-eq}
\end{align}
where $M_0$ and $\bm{H}$ define the ground state magnetization and an effective field, respectively. The first and second terms in Eq.~(\ref{ILLG-eq}) represent the traditional LLG equation~\cite{Gilbert2004}.
The inertial spin dynamics in the last term of Eq.~(\ref{ILLG-eq}) gives rise to the spin nutation~\cite{Wegrowe2012,Wegrowe2016JPCM}. { The ILLG equation signifies the fact that the dynamics of a magnetic moment shows precession with nutation at ultrafast timescales, followed by transverse damping \cite{Henk2012}. The ILLG equation has schematically been depicted in Fig.~\ref{Fig1}. }
{ A simple dimension analysis shows that }the transverse damping is characterized by a dimensionless parameter $\alpha$, and the inertial dynamics are strengthened by inertial relaxation time $\eta$.
The ILLG dynamics have been derived within the relativistic Dirac framework as well, where it shows that the Gilbert damping $\alpha$ and inertial relaxation time $\eta$ are tensors~\cite{Mondal2017Nutation}. In particular, the relativistic theory derives that the Gilbert damping dynamics is associated with the imaginary part of the susceptibility, while the inertial dynamics is given by the real part~\cite{Mondal2018JPCM}. Such findings are found to be consistent with a linear response theory of ferromagnet~\cite{Thonig_2014}. The inertial dynamics have also been derived within classical mechanics of a current loop~\cite{Giordano2020}. Eq.~(\ref{ILLG-eq}) has been applied to a single sublattice ferromagnet beyond ferromagnetic resonance (FMR), observing an additional peak due to nutation resonance~\cite{Olive2012,Olive2015,Makhfudz2020}. While the FMR peak appears at the GHz regime, the nutation resonance peak appears at the THz regime~\cite{cherkasskii2020nutation}. The ILLG equation has also been applied to antiferromagnets and ferrimagnets, and it has been predicted that the spin nutation should be better detected in antiferromagnets as it is exchange enhanced~\cite{Mondal2020nutation}.

Recently, the spin nutation resonance has been observed for ferromagnets in the experiment ~\cite{neeraj2019experimental}. Indeed, the nutation resonance peak has been seen at around 0.5 THz. Note that the experiment was performed in two-sublattice ferromagnets namely CoFeB and NiFe. For two-sublattice ferromagnet, the inter-sublattice exchange energies become important. Here, we describe the inertial effects in a two-sublattice ferromagnet coupled by the Heisenberg exchange interaction. We follow the similar procedure of Ref.~\cite{Mondal2020nutation} and derive the magnetic susceptibility. We {\it not only} consider the intra-sublattice inertial dynamics, {\it but also} the inter-sublattice dynamics. Our results suggest that there are two precession resonance peaks: one at GHz regime and another at THz regime. Similarly, two nutation peaks can also be observed, both are at the THz regime. By calculating the precession and nutation resonance frequencies, we observe that the resonance frequencies decrease with increasing intra-sublattice relaxation time, however, the scenario is different for inter-sublattice inertial dynamics.

\section{Theory of intra- and inter-sublattice inertial dynamics in two-sublattice ferromagnets}
The inertial dynamics for antiferromagnets have been introduced in Ref.~\cite{Mondal2020nutation}. 
For two-sublattice magnetic systems having magnetization  ${\bm M}_{A}$ and  ${\bm M}_{B}$, for $A$ and $B$ representing the two-sublattice, the ILLG equations of motion can be recast as
\begin{align}
\label{sublatticeA}
    \frac{\partial{\bm{M}}_{A}}{\partial t} & = -\gamma_A \left(\bm{M}_A \times \bm{H}_A\right) + \frac{\alpha_{AA}}{M_{A0}}\left(\bm{M}_A\times \frac{\partial{\bm{M}}_{A}}{\partial t}\right) + \frac{\alpha_{AB}}{M_{B0}}\left(\bm{M}_A\times \frac{\partial{\bm{M}}_{B}}{\partial t}\right) \nonumber\\
    & + \frac{\eta_{AA}}{M_{A0}}\left(\bm{M}_A\times \frac{\partial^2{\bm{M}}_{A}}{\partial t^2}\right)  + \frac{\eta_{AB}}{M_{B0}}\left(\bm{M}_A\times \frac{\partial^2{\bm{M}}_{B}}{\partial t^2}\right)\\
    \frac{\partial{\bm{M}}_{B}}{\partial t} & = -\gamma_B \left(\bm{M}_B \times \bm{H}_B\right) + \frac{\alpha_{BB}}{M_{B0}}\left(\bm{M}_B\times \frac{\partial{\bm{M}}_{B}}{\partial t}\right)  + \frac{\alpha_{BA}}{M_{A0}}\left(\bm{M}_B\times \frac{\partial{\bm{M}}_{A}}{\partial t}\right) \nonumber\\
    &+ \frac{\eta_{BB}}{M_{B0}}\left(\bm{M}_B\times \frac{\partial^2{\bm{M}}_{B}}{\partial t^2}\right) + \frac{\eta_{BA}}{M_{A0}}\left(\bm{M}_B\times \frac{\partial^2{\bm{M}}_{A}}{\partial t^2}\right)
    \label{sublatticeB}
\end{align}
In each ILLG dynamics, the first term represents the spin precession around an effective field ${\bm H}_{A/B}$. The intra- and inter-sublattice Gilbert damping dynamics have been denoted by the second and third terms, respectively. Similarly, the last two terms define inertial dynamics. While the intra-sublattice Gilbert and inertial dynamics have been weighed by $\alpha_{AA/BB}$ and $\eta_{AA/BB}$, the same for inter-sublattice dynamics are denoted by  $\alpha_{AB/BA}$ and $\eta_{AB/BA}$. From a simple dimension analysis, it is clear to show that the Gilbert damping parameters $\alpha$ are dimensionless, in contrast, the inertial relaxation times $\eta$ have a dimension of time~\cite{Ciornei2011,Mondal2017Nutation}. It is worth mentioning that the Gilbert damping $\alpha$ has been calculated for several materials within {\it ab initio} frameworks~\cite{kambersky70,kambersky76,kunes02,KunesErratum,tserkovnyak02,steiauf05,Fahnle2006,kambersky07,gilmore07,Brataas2008,Gilmore2008JAP,EbertPRL2011,Thonig_2014,Edwards_2016}, while there are also proposals to calculate the inertial relaxation time within extended breathing Fermi surface model~\cite{Fahnle2011,fahnle2013erratum,Thonig2017}.   
These ILLG equations have been contemplated to forecast the signatures of inertial dynamics in collinear antiferromagnets and ferrimagnets~\cite{Mondal2020nutation}. 

We consider that the two-sublattice ferromagnet is aligned collinear at the ground state such that ${\bm M}_{A} = M_{A0} \hat{\bm{z}}$ and ${\bm M}_{B} = M_{B0} \hat{\bm{z}}$. The ferromagnetic system is under the application of an external Zeeman field ${\bm H}_0 = H_0\hat{\bm{z}}$. Then, the free energy of the considered two-sublattice system can be considered as the sum of Zeeman, anisotropy, and exchange energies as 
\begin{align}
    \mathcal{F}\left(\bm{M}_{A}, \bm{M}_{B}\right)  & =  - H_0\left( M_{Az}+ M_{Bz}\right)  - \frac{K_A}{M^{2}_{A0}} M^2_{Az}  - \frac{K_B}{M^{2}_{B0}} M^2_{Bz} - \frac{J}{M_{A0}M_{B0}}  \bm{M}_A\cdot \bm{M}_B\,,
     \label{Free-energy}
\end{align}
where $K_A$ and $K_B$ are anisotropy energies and $J$ is the isotropic Heisenberg exchange with $J>0$ for ferromagnetic coupling. 
To calculate the linear response properties of the system, we consider that the small deviations of magnetization $\bm{m}_{A}(t)$ and $\bm{m}_{B}(t)$ with respect to the ground state are induced by the transverse external field $\bm{h}_{A}(t)$ and $\bm{h}_{B}(t)$. We calculate the effective field in the ILLG equation as the derivative of free energy in Eq.\ (\ref{Free-energy}) to the corresponding magnetization
\begin{align}
\label{HeffA}
     \bm{H}_A  = -\frac{\partial \mathcal{F}\left(\bm{M}_A, \bm{M}_B\right)}{\partial \bm{M}_A} & = \left( H_0 + \frac{2K_A}{M^{2}_{A0}}M_{Az}\right)\hat{\bm{z}}  + \frac{J}{M_{A0}M_{B0}}\bm{M}_B\nonumber\\
     & = \frac{1}{M_{A0}}\left(H_0M_{A0} + 2K_A + J\right) \hat{\bm{z}} + \frac{J}{M_{A0}M_{B0}}\bm{m}_B\,,\\
     \bm{H}_B  = -\frac{\partial \mathcal{F}\left(\bm{M}_A, \bm{M}_B\right)}{\partial \bm{M}_B}  & = \left( H_0 + \frac{2K_B}{M^{2}_{B0}}M_{Bz}\right)\hat{\bm{z}} + \frac{J}{M_{A0}M_{B0}}\bm{M}_A\nonumber\\
     & = \frac{1}{M_{B0}}\left(H_0M_{B0} + 2K_B + J\right) \hat{\bm{z}} + \frac{J}{M_{A0}M_{B0}}\bm{m}_A\,. 
     \label{HeffB}
\end{align}
%First, at the ground state, we consider that the $A$ sublattice magnetization is $\bm{M}_{A}  = M_{A0}\hat{\bm{z}}$, while the $B$ sublattice magnetization is also parallel to $A$, $\bm{M}_{B} = M_{B0}\hat{\bm{z}}$. For two sublattice ferromagnets, one can contemplate either the two sublattice magnetizations are equal ($M_{A0} = M_{B0}$) or different ($M_{A0} > M_{B0}$).
We then expand the magnetization around the ground state in small deviations, $\bm{M}_{A} = M_{A0}\hat{\bm{z}}+\bm{m}_{A} (t)$ and $\bm{M}_{B} = M_{B0}\hat{\bm{z}}+\bm{m}_{B}(t)$. Essentially, with the effective fields in Eqs.~(\ref{HeffA}) and ~(\ref{HeffB}) along with the magnetization, the linear response for sublattice A provides 
\begin{align}
  \frac{\partial{\bm{m}}_{A}}{\partial t} 
    & = -\frac{\gamma_A}{M_{A0}} \left(H_0M_{A0} + 2K_A + J\right)\left[ m_{Ay} \hat{\bm{x}} - m_{Ax} \hat{\bm{y}} \right] - \frac{\gamma_A J}{M_{B0}}\left[ m_{Bx} \hat{\bm{y}} - m_{By} \hat{\bm{x}}\right] \nonumber\\
    & - \gamma_A M_{A0} \left[ h_{Ax}\hat{\bm{y}} - h_{Ay} \hat{\bm{x}} \right] + \alpha_{AA}\left[\frac{\partial m_{Ax}}{\partial t}\hat{\bm{y}} - \frac{\partial m_{Ay}}{\partial t}\hat{\bm{x}}  \right] + \frac{\alpha_{AB}M_{A0}}{M_{B0}}\left[\frac{\partial m_{Bx}}{\partial t}\hat{\bm{y}} - \frac{\partial m_{By}}{\partial t}\hat{\bm{x}}  \right]\nonumber\\
    & + \eta_{AA} \left[\frac{\partial^2 m_{Ax}}{\partial t^2}\hat{\bm{y}} - \frac{\partial^2 m_{Ay}}{\partial t^2}\hat{\bm{x}}  \right] + \frac{\eta_{AB}M_{A0}}{M_{B0}}\left[\frac{\partial^2 m_{Bx}}{\partial t^2}\hat{\bm{y}} - \frac{\partial^2 m_{By}}{\partial t^2}\hat{\bm{x}}  \right]\,,
\end{align}
obtaining the dynamics for two components $x$ and $y$ as  
\begin{align}
    \gamma_AM_{A0} h_{Ax} & = \frac{\gamma_A}{M_{A0}} \left(H_0M_{A0} + 2K_A + J\right) m_{Ax}   - \frac{\gamma_A J}{M_{B0}} m_{Bx}  
      + \alpha_{AA}\frac{\partial m_{Ax}}{\partial t} + \frac{\alpha_{AB}M_{A0}}{M_{B0}}\frac{\partial m_{Bx}}{\partial t} \nonumber\\
    &  - \frac{\partial m_{Ay}}{\partial t}+ \eta_{AA} \frac{\partial^2 m_{Ax}}{\partial t^2}  + \frac{\eta_{AB}M_{A0}}{M_{B0}}\frac{\partial^2 m_{Bx}}{\partial t^2}\,,\\
    \gamma_AM_{A0} h_{Ay} & = \frac{\gamma_A}{M_{A0}} \left(H_0M_{A0} + 2K_A + J\right) m_{Ay}   - \frac{\gamma_A J}{M_{B0}} m_{By}  
      + \alpha_{AA}\frac{\partial m_{Ay}}{\partial t} + \frac{\alpha_{AB}M_{A0}}{M_{B0}}\frac{\partial m_{By}}{\partial t} \nonumber\\
    &  + \frac{\partial m_{Ax}}{\partial t}+ \eta_{AA} \frac{\partial^2 m_{Ay}}{\partial t^2}  + \frac{\eta_{AB}M_{A0}}{M_{B0}}\frac{\partial^2 m_{By}}{\partial t^2}\,.
\end{align}
In the circular basis defined by $m_{A\pm} = m_{Ax} \pm {\rm i}m_{Ay}$ and $h_{A\pm} = h_{Ax} \pm {\rm i}h_{Ay}$, the equations can be put together 
\begin{align}
    \gamma_AM_{A0} h_{A\pm} & = \frac{\gamma_A}{M_{A0}} \left(H_0M_{A0} + 2K_A + J\right) m_{A\pm}   - \frac{\gamma_A J}{M_{B0}} m_{B\pm}  
      + \alpha_{AA}\frac{\partial m_{A\pm}}{\partial t} + \frac{\alpha_{AB}M_{A0}}{M_{B0}}\frac{\partial m_{B\pm}}{\partial t} \nonumber\\
    &  \pm {\rm i}\frac{\partial m_{A\mp}}{\partial t}+ \eta_{AA} \frac{\partial^2 m_{A\pm}}{\partial t^2}  + \frac{\eta_{AB}M_{A0}}{M_{B0}}\frac{\partial^2 m_{B\pm}}{\partial t^2}\,.
    \label{h_A}
\end{align}
Similarly, one can calculate the linear response of the sublattice B in the circular basis defined by $m_{B\pm} = m_{Bx} \pm {\rm i}m_{By}$ and $h_{B\pm} = h_{Bx} \pm {\rm i}h_{By}$ as 
\begin{align}
    \gamma_B M_{B0} h_{B\pm} & = \frac{\gamma_B}{M_{B0}} \left(H_0M_{B0} + 2K_B + J\right) m_{B\pm}   - \frac{\gamma_B J}{M_{A0}} m_{A\pm}  
      + \alpha_{BB}\frac{\partial m_{B\pm}}{\partial t} + \frac{\alpha_{BA}M_{B0}}{M_{A0}}\frac{\partial m_{A\pm}}{\partial t} \nonumber\\
    &  \pm {\rm i}\frac{\partial m_{B\mp}}{\partial t}+ \eta_{BB} \frac{\partial^2 m_{B\pm}}{\partial t^2}  + \frac{\eta_{BA}M_{B0}}{M_{A0}}\frac{\partial^2 m_{A\pm}}{\partial t^2}\,.
    \label{h_B}
\end{align}
We define the response functions $m_{A\pm},m_{B\pm},h_{A\pm},h_{B\pm}  \propto e^{\pm {\rm i} \omega t}$ and  $\Omega_A = \frac{\gamma_A}{M_{A0}} \left(H_0M_{A0} + 2K_A + J\right)$ and $\Omega_B = \frac{\gamma_B}{M_{B0}} \left(H_0M_{B0} + 2K_B + J\right)$. %The two equations can be written as
%\begin{align}
%    h_{A\pm} & = \frac{1}{\Gamma_{AA}}\left(\Omega_A \pm {\rm i}\omega \alpha_{AA} - \omega^2 \eta_{AA} - \omega\right) m_{A\pm}   - \frac{1}{\Gamma_{AB}}\left(\frac{\gamma_A J}{M_{A0}} 
%       \mp {\rm i}\omega \alpha_{AB}    + \omega^2 \eta_{AB}\right) m_{B\pm}\nonumber\\
%     h_{B\pm} & = \frac{1}{\Gamma_{BB}} \left(\Omega_B \pm {\rm i}\omega \alpha_{BB} - \omega^2 \eta_{BB} - \omega\right) m_{B\pm}   - \frac{1}{\Gamma_{BA}}\left(\frac{\gamma_B J}{M_{B0}} 
%       \mp {\rm i}\omega \alpha_{BA}    + \omega^2 \eta_{BA}\right) m_{A\pm}
%\end{align}
To simplify the expressions, we introduce the following: $\Gamma_{AA} = \gamma_A M_{A0}$, $\Gamma_{BB} = \gamma_B M_{B0}$, $\Gamma_{AB} = \gamma_A M_{B0}$ and $\Gamma_{BA} = \gamma_B M_{A0}$ such that $\Gamma_{AA}\Gamma_{BB} =\Gamma_{AB}\Gamma_{BA} $. The linear response Eqs.~(\ref{h_A}) and (\ref{h_B}) can be written in a matrix formalism 
\begin{align}
    & \begin{pmatrix}
      h_{A\pm}\\
       h_{B\pm}
    \end{pmatrix} \nonumber\\
    & = \begin{pmatrix}
      \dfrac{1}{ \Gamma_{AA}}\left(\Omega_A \pm {\rm i}\omega \alpha_{AA} - \omega^2 \eta_{AA} - \omega\right) & -\dfrac{1}{\Gamma_{AB}} \left(\dfrac{\gamma_A J}{M_{A0}} 
       \mp {\rm i}\omega \alpha_{AB}    + \omega^2 \eta_{AB}\right)\\
     -\dfrac{1}{\Gamma_{BA}} \left(\dfrac{\gamma_B J}{M_{B0}} 
       \mp {\rm i}\omega \alpha_{BA}    + \omega^2 \eta_{BA}\right) &  \dfrac{1}{ \Gamma_{BB}}\left(\Omega_B \pm {\rm i}\omega \alpha_{BB} - \omega^2 \eta_{BB} - \omega\right)
    \end{pmatrix}\begin{pmatrix}
      m_{A\pm}\\
      m_{B\pm}
    \end{pmatrix}\,.
\end{align}
For finding the susceptibility, we recall $\bm{m}_{\pm} = \chi_{\pm}\cdot {\bm h}_{\pm}$ such that the susceptibility matrix derives as
\begin{align}
   \chi^{AB}_\pm & = \dfrac{1}{\mathcal{D}_\pm}\begin{pmatrix}
      \dfrac{1}{ \Gamma_{BB}}\left(\Omega_B \pm {\rm i}\omega \alpha_{BB} - \omega^2 \eta_{BB} - \omega\right)  & \dfrac{1}{\Gamma_{BA}} \left(\dfrac{\gamma_B J}{M_{B0}} 
       \mp {\rm i}\omega \alpha_{BA}    + \omega^2 \eta_{BA}\right)\\
    \dfrac{1}{\Gamma_{AB}} \left(\dfrac{\gamma_A J}{M_{A0}} 
       \mp {\rm i}\omega \alpha_{AB}    + \omega^2 \eta_{AB}\right)&  \dfrac{1}{ \Gamma_{AA}}\left(\Omega_A \pm {\rm i}\omega \alpha_{AA} - \omega^2 \eta_{AA} - \omega\right)
    \end{pmatrix}\,,
\end{align}
where the determinant is expressed as \begin{align}
   \mathcal{D}_\pm & = \frac{1}{\Gamma_{AA}\Gamma_{BB}} \left(\Omega_A \pm {\rm i}\omega \alpha_{AA} - \omega^2 \eta_{AA} - \omega\right)\left(\Omega_B \pm {\rm i}\omega \alpha_{BB} - \omega^2 \eta_{BB} - \omega\right)  \nonumber\\
   &  - \frac{1}{\Gamma_{AB}\Gamma_{BA}} \left(\frac{\gamma_A J}{M_{A0}} \mp {\rm i}\omega \alpha_{AB}    + \omega^2 \eta_{AB}\right)\left(\frac{\gamma_B J}{M_{B0}} \mp {\rm i}\omega \alpha_{BA}    + \omega^2 \eta_{BA}\right)\,.
\end{align}
Note that the intra-sublattice dynamical parameters enter in the diagonal elements of the susceptibility matrix, however, the inter-sublattice dynamics are reflected in the off-diagonal elements. Such a susceptibility matrix has been obtained with intra- and inter-sublattice Gilbert damping dynamics for antiferromagnets~\cite{Kamra2018}.   
To find the resonance frequencies, one has to solve the equation setting $\mathcal{D}_{\pm} = 0$. 
Therefore, a fourth-order equation in frequency is obtained 
\begin{align}
    \mathbb{A}_\pm \omega^4 + \mathbb{B}_\pm \omega^3 + \mathbb{C}_\pm \omega^2 + \mathbb{D}_\pm \omega + 
    \mathbb{E}_\pm & = 0\,,
    \label{4th-order-Eq}
\end{align}
with the following coefficients
\begin{align}
    \mathbb{A}_\pm & = \eta_{AA}\eta_{BB} - \eta_
    {AB}\eta_{BA}\,,\\
    \mathbb{B}_\pm & =(\eta_{AA} + \eta_{BB}) \mp {\rm i} \left(\alpha_{AA}\eta_{BB} + \alpha_{BB}\eta_{AA}\right) \pm {\rm i}\left(\alpha_{AB}\eta_{BA}+ \alpha_{BA}\eta_{AB}\right)\,,\\
    \mathbb{C}_\pm & =1 \mp {\rm i} \left(\alpha_{AA} + \alpha_{BB}\right)- \left(\Omega_A\eta_{BB} + \Omega_B\eta_{AA}\right) - \alpha_{AA}\alpha_{BB}\nonumber\\
    & -\left(\frac{\gamma_A}{M_{A0}}\eta_{BA} + \frac{\gamma_B}{M_{B0}}\eta_{AB}\right)J-\alpha_{AB}\alpha_{BA}\,,\\
    \mathbb{D}_\pm & =-\left(\Omega_A + \Omega_B\right)\pm {\rm i} \left(\Omega_A\alpha_{BB} +\Omega_B\alpha_{AA} \right) \pm{\rm i} \left(\frac{\gamma_A}{M_{A0}}\alpha_{BA} + \frac{\gamma_B}{M_{B0}}\alpha_{AB}\right)J\,,\\
    \mathbb{E}_\pm & =\Omega_A \Omega_B -\frac{\gamma_A\gamma_B}{M_{A0}M_{B0}}J^2\,.
\end{align}
%\begin{align}
%    & \left[\eta_{AA}\eta_{BB} - \eta_
%    {AB}\eta_{BA}\right] \omega^4 + \left[(\eta_{AA} + \eta_{BB}) \mp {\rm i} \left(\alpha_{AA}\eta_{BB} + \alpha_{BB}\eta_{AA}\right) \pm {\rm i}\left(\alpha_{AB}\eta_{BA}+ \alpha_{BA}\eta_{AB}\right) \right]\omega^3\nonumber\nonumber\\
%    & + \left[1 \mp {\rm i} \left(\alpha_{AA} + \alpha_{BB}\right)- \left(\Omega_A\eta_{BB} + \Omega_B\eta_{AA}\right) - \alpha_{AA}\alpha_{BB}-\left(\frac{\gamma_A}{M_{A0}}\eta_{BA} + \frac{\gamma_B}{M_{B0}}\eta_{AB}\right)J-\alpha_{AB}\alpha_{BA} \right]\omega^2\nonumber\\
%    & + \left[-\left(\Omega_A + \Omega_B\right)\pm {\rm i} \left(\Omega_A\alpha_{BB} +\Omega_B\alpha_{AA} \right) \pm{\rm i} \left(\frac{\gamma_A}{M_{A0}}\alpha_{BA} + \frac{\gamma_B}{M_{B0}}\alpha_{AB}\right)J \right]\omega \nonumber\\
%    & + \left(\Omega_A \Omega_B -\frac{\gamma_A\gamma_B}{M_{A0}M_{B0}}J^2\right) = 0 
%\end{align}
The analytical solution of the above-mentioned equation is very cumbersome. Therefore, we adopt the numerical techniques for solving Eq.~(\ref{4th-order-Eq}). 
The solution of the above equation results in four frequencies, two of them correspond to the precession resonance ($\omega_{\rm p}$) of each sublattice and the other two belong to the nutation resonance ($\omega_{\rm n}$). The real and imaginary parts of the resonance frequency are denoted by ${\tt Re}$ and ${\tt Im}$, respectively. For example, the precession resonance frequencies are $\omega_{\rm p} = {\tt Re}\left(\omega_{\rm p}\right) + {\rm i}{\tt Im}\left(\omega_{\rm p}\right)$, while the nutation resonance frequencies are $\omega_{\rm n} = {\tt Re}\left(\omega_{\rm n}\right) + {\rm i}{\tt Im}\left(\omega_{\rm n}\right)$.
Comparing Eq.~(\ref{4th-order-Eq}), a similar equation has been obtained for antiferromagnets and ferrimagnets~\cite{Mondal2020nutation}, however, without the inter-sublattice inertial dynamics. We mention that the inter-sublattice Gilbert damping dynamics have extensively been discussed~\cite{Kamra2018,Yuan_2019}. Therefore, we will not consider in the following discussions. In particular, we allow $\alpha_{AB} = \alpha_{BA} = 0$, and calculate the inertial effects on precession and nutation resonances.   

\section{Numerical results}
To calculate the resonace frequencies, we numerically solve the Eq.~(\ref{4th-order-Eq}) for two-sublattice ferromagnets having same magnetic moments in each sublattice i.e., $M_{A0} = M_{B0}$.
We use the following parameters: $\gamma_A = \gamma_B = 1.76\times 10^{11}$ T$^{-1}$-s$^{-1}$, $J = 10^{-21}$ J, $K_A = K_B = 10^{-23}$ J, $\alpha_{AA} = \alpha_{BB} =  \alpha =  0.05$, $\alpha_{AB} = \alpha_{BA} = 0$. The considered exchange and anisotropy energies have similar order of magnitude as typical ferromagnets e.g., Fe \cite{Pajda2001}. The chosen Gilbert damping $\alpha = 0.05$ is within the {\it ab initio} reported values~\cite{EbertPRL2011}. For inertial relaxation times, even though, the {\it ab initio} calculation suggests about fs timescales for transition metals~\cite{Thonig2017}, the recent experiment predicts it to be a higher value up to several hundreds of fs~\cite{neeraj2019experimental}. Therefore, in what follows, we have considered the inertial relaxation times ranging from fs to ps.        

\subsection{Intra-sublattice inertial dynamics}
\begin{figure}[hbt!]
    \centering
    \includegraphics[scale = 0.22]{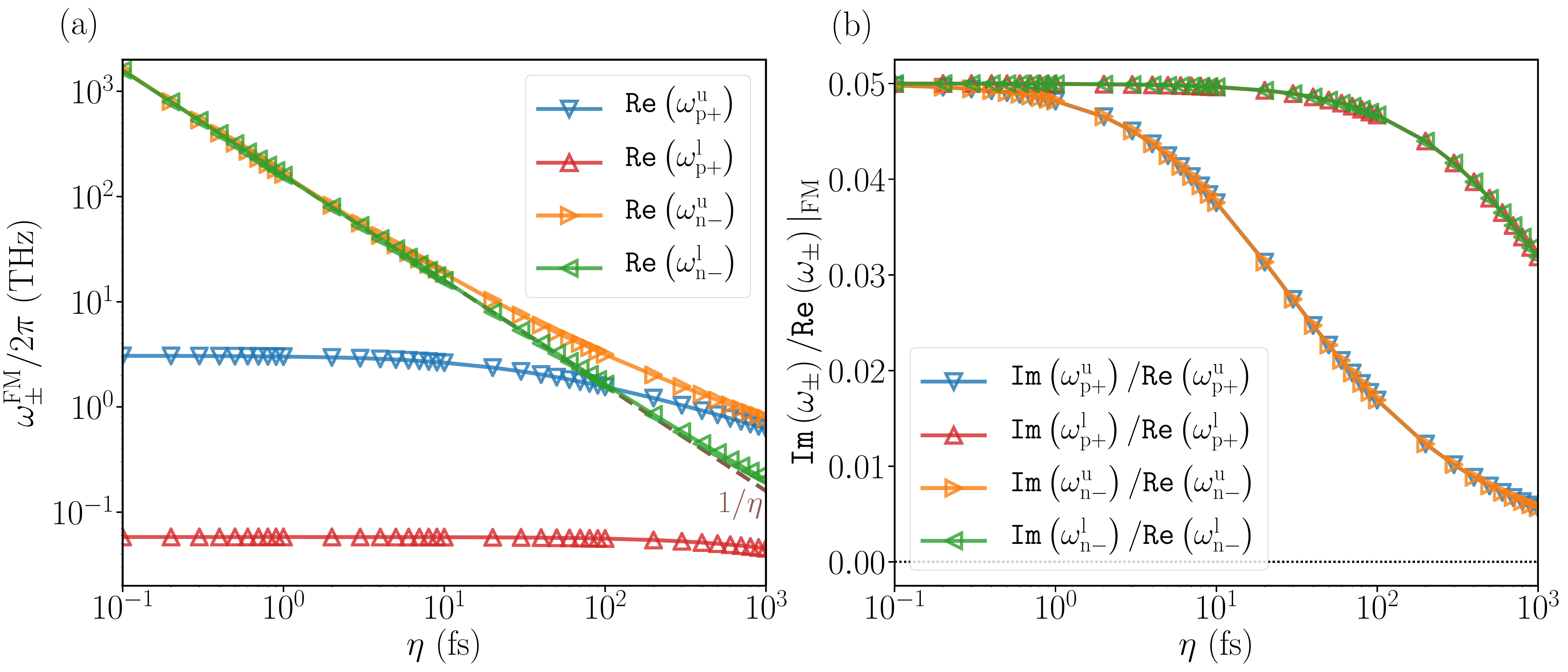}
    \caption{The calculated resonance frequencies  as a function of intra-sublattice inertial relaxation time for two-sublattice ferromagnets using $M_{A0} = M_{B0} = 2\mu_B$. (a) The precession and nutation resonance frequencies and (b) the effective Gilbert damping have been plotted.}
    \label{intra-sublattice-Fig1}
\end{figure}
\begin{figure}[hbt!]
    \centering
    \includegraphics[scale = 0.9]{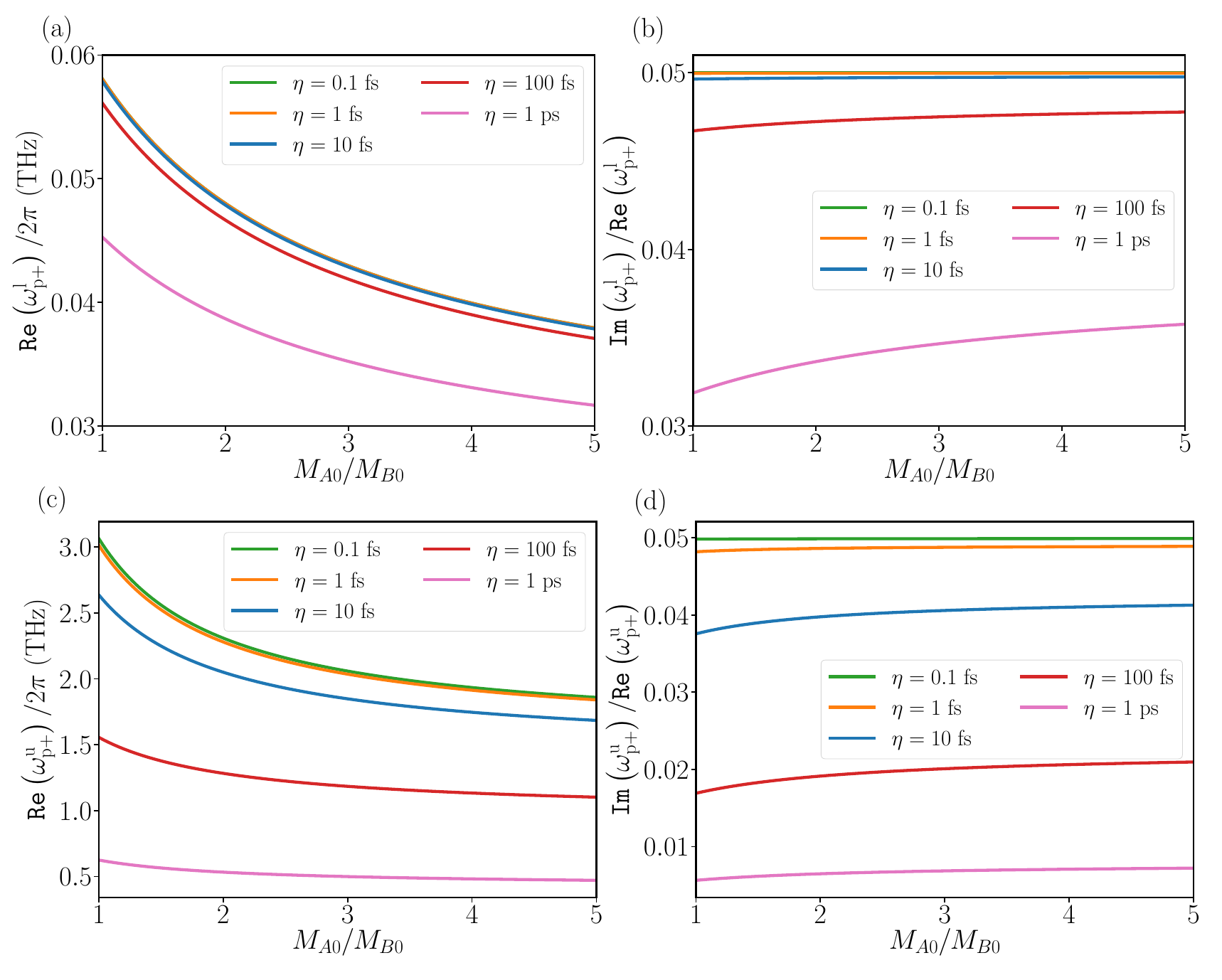}
    \caption{The calculated precession resonance frequencies as a function of $M_{A0}/M_{B0}$ for two-sublattice ferromagnets, at several intra-sublattice inertial relaxation times. (a) The real part of the lower precession resonance frequencies,  (b) the effective damping of lower resonance mode, (c) the real part of the upper precession resonance frequencies,  (d) the effective damping of upper resonance mode, has been plotted.}
    \label{intra-sublattice-Fig2}
\end{figure}
 To focus on the intra-sublattice inertial dynamics, we set the inter-sublattice relaxation time to zero i.e., $\eta_{AB} = \eta_{BA}$ = 0, keeping the same inertial relaxation time in two-sublattice $\eta_{AA} = \eta_{BB} = \eta$. With this set of specifications, the calculated frequencies have been shown in Fig.~\ref{intra-sublattice-Fig1}. One can see that there exist two precession resonance frequencies (positive) and the corresponding two nutation resonance frequencies (negative). We denote these two positive precession frequencies as $\omega_{\rm p+}^{\rm u}$ and $\omega_{\rm p+}^{\rm l}$, while the two negative nutation frequencies are $\omega_{\rm n-}^{\rm u}$ and $\omega_{\rm n-}^{\rm l}$. The superscripts ``u'' and ``l'' denote the upper and lower frequencies, respectively. These results are in contrast with the observation in antiferromagnets or ferrimagnets, where one positive and one negative precession (and nutation) frequencies are expected \cite{Mondal2020nutation}. { Nevertheless, the quantitative comparison of the calculated frequencies agrees with those of the ferrimagnets, where the upper (THz), and lower (GHz) frequency precession resonances are called an exchange and ferromagnetic modes, respectively~\cite{Mondal2020nutation,Schlickeiser2012}.  Similar to antiferromagnets and ferrimagnets~\cite{Mondal2020nutation}, the resonance frequencies decrease with the intra-sublattice inertial relaxation time in the case of two-sublattice ferromagnets. Especially, the lower nutation resonance frequency scales with $1/\eta$, while the upper one shows deviation from $1/\eta$ at higher relaxation times. This deviation from $1/\eta$ has been noticed in two nutation modes for antiferromagnets and ferrimagnets~\cite{Mondal2020nutation}. An interesting feature is that the precession and nutation frequencies cross each other at certain inertial relaxation times in ferromagnets. Such crossing was not observed in antiferromagnets and ferrimagnets~\cite{Mondal2020nutation}. The crossing happens especially with the upper precession mode with lower nutation mode as seen in Fig.~\ref{intra-sublattice-Fig1}(a). However, we note that crossing of these two modes have positive and negative frequencies, meaning that the upper precession mode ($\omega_{\rm p+}^{\rm u}$) has a positive rotational sense, however, the lower nutation mode ($\omega_{\rm n-}^{\rm l}$) has the opposite rotational sense in circular basis. 

The inertial dynamics affect the effective Gilbert damping in a system. This has been demonstrated in Fig.~\ref{intra-sublattice-Fig1}(b) for two-sublattice ferromagnet by the ratio of imaginary and real parts of the calculated frequencies. We have used the same Gilbert damping for both the sublattices $\alpha \sim 0.05$ and therefore, the effective damping remains the same at smaller inertial relaxation times. However, the effective damping decreases with increased relaxation times, a fact that is consistent with the results of antiferromagnets \cite{Mondal2020nutation}. It is observed that the decrease in effective damping is exactly the same for precession and corresponding nutation modes. Moreover, the upper precession mode is influenced strongly, which has already been observed for ferrimagnets \cite{Mondal2020nutation}.

Next, we calculate the influence of different sublattice magnetic moment ($M_{A0} \neq M_{B0}$) on inertial dynamics. In particular, we compute the precession resonance frequencies as a function of the ratio of magnetic moments ($M_{A0}/M_{B0}$), at several inertial relaxation times in Fig.~\ref{intra-sublattice-Fig2}. We observe that the resonance frequencies decrease with increasing difference in the magnetic moments. Such reduction is less visible in case of lower precession frequencies e.g., Fig.~\ref{intra-sublattice-Fig2}(a), however, more prominent in upper precession frequencies in Fig.~\ref{intra-sublattice-Fig2}(c). However, the difference of frequencies calculated at several relaxation times are similar for $M_{A0} = M_{B0}$ and $M_{A0} \neq M_{B0}$.
The latter suggests that the inertial dynamics do not get quantitatively influenced by the same or different sublattice magnetic moments. A similar conclusion can also be made from the computation of effective damping in Figs.~\ref{intra-sublattice-Fig2}(b) and \ref{intra-sublattice-Fig2}(d). The effective damping for the upper and lower precession modes remains almost constant (with a very small positive slope) for a higher ratio of $M_{A0} / M_{B0}$.      

\subsection{Inter-sublattice inertial dynamics}
\begin{figure}[hbt!]
    \centering
    \includegraphics[scale = 0.22]{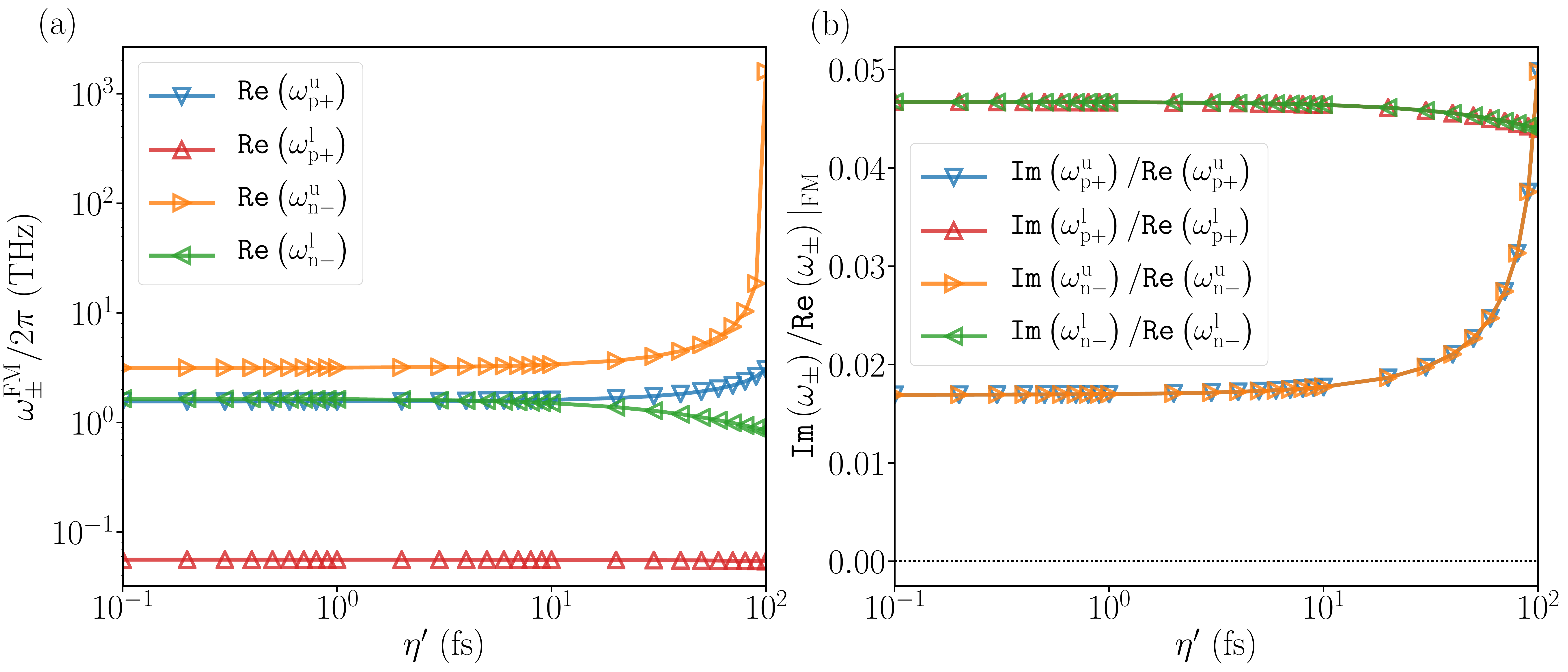}
    \caption{The calculated resonance frequencies  as a function of inter-sublattice inertial relaxation time for two-sublattice ferromagnets using $M_{A0} = M_{B0} = 2\mu_B$. The intra-sublattice inertial relaxation time was kept constant $\eta = 100$ fs.  (a) The precession and nutation resonance frequencies and (b) the effective Gilbert damping have been plotted.}
    \label{inter-sublattice-Fig3}
\end{figure}
To investigate the inter-sublattice inertial dynamics, we set the intra-sublattice relaxation time as $\eta_{AA} = \eta_{BB} = \eta = 100$ fs. Such a relaxation time is lower than the experimental findings in two-sublattice ferromagnets~\cite{neeraj2019experimental}. In fact, the direct comparison of Eq.~(\ref{sublatticeA}) with the Eq.~(2) of Ref.~\cite{neeraj2019experimental} provides $\eta \sim \alpha \tau$. With the experimental findings for CoFeB, $\alpha = 0.0044$ and $\tau = 72$ ps (see Table 1 in Ref.~\cite{neeraj2019experimental}), we calculate $\eta = 316$ fs. We compute the effect of inter-sublattice inertial dynamics as a function of $\eta_{AB} = \eta_{BA} = \eta^\prime$ in Fig.~\ref{inter-sublattice-Fig3} considering $\eta^\prime < \eta$.   
As we mentioned earlier, the overlapping of precession ($\omega_{\rm p+}^{\rm u}$) and nutation ($\omega_{\rm n-}^{\rm l}$) frequencies at the intra-sublattice relaxation time $\eta = 100$ fs can be seen. 
We observe that the upper precession resonance frequency ($\omega_{\rm p+}^{\rm u}$) increases, while the lower one ($\omega_{\rm p+}^{\rm l}$) decreases very small with inter-sublattice relaxation times. A similar conclusion can be made for nutation frequencies. This is in contrast to the observation of intra-sublattice inertial dynamics as discussed above. A divergence in the upper nutation frequency can be noticed at the limit $\eta^\prime \rightarrow \eta$. Such divergence can be explained through the coefficient $\mathbb{A}$ in Eq.~(\ref{4th-order-Eq}). At the limit $\eta^\prime \rightarrow \eta$, the coefficient of fourth power in frequency becomes $\mathbb{A} = \eta_{AA}\eta_{BB} - \eta_{AB}\eta_{BA} = \eta^2 - \eta^{\prime 2} \rightarrow 0$, which brings the fourth-order equation into an effective third-order equation in frequency. 

A similar observation can also be concluded from the calculation of effective damping in Fig.~\ref{inter-sublattice-Fig3}(b). Similar to the intra-sublattice inertial dynamics, the effective damping of the precession and corresponding nutation mode behaves exactly the same for the inter-sublattice inertial dynamics. We observe that the damping of upper precession and nutation modes increases with inter-sublattice inertial relaxation time, however, it is the opposite for lower precession and nutation modes. Therefore, we conclude that the effect of intra- and inter-sublattice inertial dynamics are contrasting.

\section{Conclusions}
To conclude, we have incorporated the intra- and inter-sublattice inertial dynamics within the LLG equation of motion and calculated the FMR resonance for two-sublattice ferromagnets. To this end, we first derive the magnetic susceptibility that is a tensor. To calculate the resonance frequencies, we find the poles of the susceptibility. Without the inertial dynamics, there exist two precession modes in a typical two-sublattice ferromagnet. The introduction of inertial dynamics shows two nutation resonance frequencies corresponding to the precession modes. We note that these precession and nutation resonances can be excited by right and left circularly polarised pulses, respectively, and vice-versa within a circular basis. The precession and nutation frequencies decrease with the intra-sublattice relaxation time as also has been seen in the case of antiferromagnets in previous work~\cite{Mondal2020nutation}. However, at certain relaxation times, the precession and nutation frequencies overlap with each other. Note that these overlapping precession and nutation frequencies have opposite rotational sense in circular basis, thus, they can be neatly realised in the experiments. The inter-sublattice inertial dynamics increase the resonance frequencies and effective damping for upper precession mode, however, have opposite effect on lower precession mode in two-sublattice ferromagnets.

\section{Acknowledgments}
The author acknowledges Levente R\'ozsa and Ulrich Nowak for valuable discussions and the Swedish Research Council (VR 2019-06313) for research funding.

\bibliographystyle{iopart-num}
%\bibliography{References}

%\end{document}
\providecommand{\newblock}{}

\end{document}